\documentstyle{article}

\title{\sc QUANTUM STATE OF WORMHOLES AND TOPOLOGICAL ARROW OF TIME}
\author{ Pedro F. Gonz\'alez-D\'{\i}az.\\
Instituto de Matem\'aticas y F\'{\i}sica Fundamental\\
Consejo Superior de Investigaciones Cientificas\\
Serrano 121, 28006 Madrid (SPAIN)\\
}
\date{May 21, 1994}
\begin{document}
\maketitle
\large
\setlength{\baselineskip}{0.5cm}
\vspace{3cm}

\begin{abstract}

This paper studies the time-symmetry problem in quantum gravity. The issue
depends critically on the choice of the quantum state and has been considered
in this paper by restricting to the case of quantum wormholes. It is seen that
pure states represented by a wave functional are time symmetric. However, a
maximal analytic extension of the wormhole manifold is found that corresponds
to a mixed state describable by a nondegenerate density-matrix functional that
involves an extra quantum uncertainty for the three-metric, and is free from
the
divergences encountered so far in statistical states formulated in quantum
gravity. It is then argued that, relative to one
asymptotic region, the statistical quantum state of single Euclidean wormholes
in semiclassical approximation is time-asymmetric and
gives rise to a $topological$ arrow of time which will reflect in the set of
all quantum fields at low energies of the asymptotically flat region.

\end{abstract}


\section{Introduction}

There has been some controversy concerning whether quantum gravity is time
symmetric [1-4].
The debate has been most emphasized around the ideas of Hawking and Penrose.
The case for a time-symmetric quantum gravity was first presented by Hawking
who
suggested [1], by considering the state of thermal equilibrium of a black hole
in a large container with perfectly reflecting walls, that since the essential
physical theories involved are time-symmetric, the equilibrium state ought to
be time-symmetric too. Penrose has argued [2] against this conclusion, and
suggested in turn his own which favours a time-asymmetric quantum gravity, as
expressed in terms of his general Weyl curvature hypothesis [4]. However,
before worrying about time-symmetry in quantum gravity, one should make
it clear that the very concept of time is far from being well defined in
quantum gravity. Actually, such a concept is under active investigation
and some proposals have been advanced in recent years [5].

Although originally
aimed at showing that white holes ought to be physically indistinguishable from
black holes [1], Hawking gave more recently [3] the idea of a time-symmetric
quantum gravity a new more general form. By contrast to how the Penrose's
universal boundary condition associated with the Weyl curvature
hypothesis, implied a time-asymmetric quantum gravity, the no boundary
proposal [6] suggested by Hawking [7] enabled him to show [3] that the quantum
state of the universe
must be $CTP$ and $T$ invariant simultaneously. This is not the case, however,
for
the other nowadays major competitor proposal for the initial conditions
advanced
by Vilenkin [8].

On the other hand, in any theory of quantum gravity one should allow baby
universes to branch off from our nearly flat region of spacetime through
wormholes [9].
It is the aim of this work to investigate
whether the quantum state of Euclidean wormholes is $T$, $CT$ and $CTP$
invariant and, in particular, what
could be the effects of these wormholes in the $T$-invariance properties of
the whole set of the resulting effective
quantum field theories at low energy in the asymptotic region.

\section{THE QUANTUM STATE OF WORMHOLES}
\setcounter{equation}{0}

The question that arises is, what is the quantum state of a wormhole or little
baby universe?. It was first considered [10] that the quantum state of a
wormhole
should be given by a wave functional of the three-metric, $h_{ij}$, and
matter field configurations, $\phi$, on the inner three-manifold, which is
given by a path integral
\begin{equation}\Psi[h_{ij},\phi_{0}] = \int_{C_{\Psi}}Dg_{\mu\nu}\delta\phi
e^{-I[g_{\mu\nu},\phi]},\end{equation}
where $I$ is the Euclidean wormhole action and the integration is over the
class $C_{\Psi}$ of asymptotically flat Euclidean four-geometries and
asymptotically vanishing matter field configurations which match the prescribed
data
on an inner three-surface $S$ with topology $S^{3}$ which divides the whole
manifold into two disconnected parts. Thus, the wormhole is in a pure state.
The condition that matter fields should vanish asymptotically arises from
the boundary conditions [11]. For asymptotically Euclidean four-geometries,
the Euclidean action expressed in Hamiltonian form, $\tilde{I}_{H}$,
amounted with an additional surface term at time $\tau\rightarrow\infty$
to make the action finite, must be invariant under reparametrizations
on the three-surface [12]. One should then remove the resulting gauge
freedom by imposing that the induced variation of the action,
$\delta\tilde{I}_{H}$, be exactly zero. This, in turns implies that,
if no cosmological term is present and the gravitational part of the
Hamiltonian constraint goes to zero as $\tau\rightarrow\infty$, the
matter field potential should strictly vanish asymptotically, leaving
no matter excitation on the large region.

More recently, however, it has been suggested [13] that the most general
quantum
state of wormholes need not be in a pure state. Actually, because of lack of
any
precise knowledge about the baby universe system, a statistical mixed state
should represent better these wormholes, and therefore, a non-factorizable
density matrix functional $\rho$ was proposed [13], rather than $\Psi$, to
describe the quantum state of wormholes. Thus, this state is made most general
by considering a path integral
\begin{equation}\rho[h_{ij},\phi_{0};h'_{ij},\phi '_{0}] =
\int_{C_{\rho}}Dg_{\mu\nu}\delta\phi e^{-I[g_{\mu\nu},\phi]},\end{equation}
where the integration is over a class $C_{\rho}$ of asymptotically flat
four-geometries
and asymptotically vanishing matter field configurations which match the
prescribed
data $[h_{ij},\phi_{0}]$ on its inner three-surface $S$, and the orientation
reverse of the configuration $[h'_{ij},\phi '_{0}]$ on its copy three-manifold
$S'$,
$S$ and $S'$ together forming the inner boundary of each four-geometry in the
path integral. The proposal includes those contributions coming from connected
four-geometries joining $S$ and $S'$, where the inner boundary does not divide
the manifold.

Both the wave function and the density matrix for wormholes have been worked
out
explicitely from the Wheeler DeWitt equation, subjected to suitable boundary
conditions. For the most general scalar-field wormhole, they are given in terms
of harmonic
oscillator wave functions [10]
\begin{equation}\Psi[a,f] =
H_{n}(a)e^{-\frac{1}{2}a^{2}}H_{m}(f)e^{-\frac{1}{2}f^{2}}\end{equation}
(in which for the sake of simplicity in the formulation only the lowest
homogeneous harmonic mode in the expansion of the scalar field has been
considered) and the propagator [13]
\begin{equation}K_{0}[a,f,0;a',f',\tau]=\sum_{m}\sum_{n}\Psi_{mn}[a,f]\Psi_{mn}[a',f']e^{-\epsilon_{mn}\tau}, \end{equation}
where $a$ and $a'$ are the radius of the inner three-spheres $S$ and $S'$, $f$
and $f'$ are the mode coefficients of the scalar harmonics in terms of which
we expand the fields $\phi$ and $\phi '$, $H(x)$ denotes Hermite polynomials,
$\tau$ measures the Euclidean time separation between any two three-surfaces
$S$ and $S'$
for which the arguments of the functional are given,
and $\epsilon_{mn}$ is expressed in terms of the energy levels of the
matter fields, $E^{f}_{m}$, and gravitational field, $E^{a}_{n}$, as
\begin{equation}\epsilon_{mn} = E^{f}_{m} - E^{a}_{n}.\end{equation}
For observers living inside the Lorentzian light cone of one asymptotic region,
$S$ and $S'$ may have any Euclidean time separation and, therefore, the proper
density matrix operator would be obtained by integrating $K_{0}$ over all
positive values of $\tau$ between 0 and $\infty$. In this case, however, one
would obtain a divergent result if $\epsilon_{mn}$ is allowed to vanish
(Note that $\epsilon_{mn}=0$ would actually correspond to the ground-state
wave function where one should not need to integrate over $\tau$ as the two
resulting submanifolds are now disconnected)
or
to take on negative values as one should actually do because $\epsilon_{mn}$
is a discontinuous quantity. Clearly, divergences come about because both
$\epsilon_{mn}$ and $\tau$ enter the exponent in $K_{0}$ linearly and there is
no nonzero vacuum zero-point energy preventing $\epsilon_{mn}$ to vanish.

If the wave functional would be evaluated by a path integration using
semiclassical approximation and higher, inhomogeneous field modes
were excited, then generally there will be a saddle point for each
configuration. The effect of allowing excitation of higher,
homogeneous modes on the density matrix would manifest by the
presence of a product $\prod H_{m_{j}}(f_{j})e^{-\frac{1}{2}f_{j}^{2}}$,
instead of the single $H_{m}(f)e^{-\frac{1}{2}f^{2}}$, in the
functionals entering (2.4), so as by a more complicate exponent in
the propagator, $\epsilon_{m_{j}n}\tau=(\sum E_{m_{j}}^{f_{j}}-E_{n}^{a})\tau$,
instead of (2.5), with $\tau$ also varying between $0$ and $\infty$.
Furthermore, one should sum over each $m_{j}$ in the propagator.
These modifications do not change however the conclusions of the discussion
made for the case when only wormholes with the lowest, homogeneous
scalar field mode $j=1$ excited, are considered.

It can be argued [14] that the analogy between the standard theory
and the gravitational theory used to formulate the propagator (2.4) is
incomplete. This can be seen by considering that in the current derivation of
the
ground-state wave functional from the standard propagator
\[\sum_{n}\sum_{m}\Psi_{mn}(x)\Psi_{mn}(x')\exp[i\epsilon_{mn}(t-t')]\]
one currently makes a rotation $t\rightarrow -i\tau$ and takes the limit as
$\tau\rightarrow -\infty$ [6]. Then, only the term $\epsilon_{mn}=0$ would
survive if all other $\epsilon_{mn}>0$, but the functional does diverge for all
those eigenenergies which are negative, $\epsilon_{mn}<0$. If we Wick rotate
$t\rightarrow +i\tau$ [14], the situation would become just the opposite, with
the divergences arising then from all $\epsilon_{mn}>0$ if we take the limit
$\tau\rightarrow -\infty$, or from $\epsilon_{mn}<0$ in the limit
$\tau\rightarrow +\infty$. In the next section
we will devise a method for obtaining a consistent propagator in superspace
which is free from such
divergences.

\section{QUANTUM STATE AND SPHALERONS}
\setcounter{equation}{0}

Many field theories have a degenerate vacuum structure showing more than one
potential minimum. Such a complicate vacuum structure makes it possible the
occurrence at zero energy of quantum transitions describable as instantons
between states lying in the vecinity of different vacua. Instantons are
localized objects which correspond to solutions of the Euclidean equations
of motion with finite action [15]. On the other hand, for nonzero energies,
transitions between distinct vacua may also occur classically by means of
sphalerons. Whereas instantons tunnel from one potential minimum to another
by going below the barrier, sphalerons do the transit over the barrier.
Sphalerons correspond to the top of this barrier and are unstable classical
solutions to the field equations which are static and localized in space as
well [16,17].

In this section, we shall first review the arguments that make it possible
the existence
of new sphaleron-like
transitions which are classically forbidden though they may still take place
in the quantum-mechanical realm [18], relating then these quantum-sphaleron
transitions with the admissible quantum states of wormholes.
Such transitions would typically pertain
to nonlinear systems showing bifurcations phenomena.
In order to see how
these nonclassical transitions may appear, let us consider a theory with
the Euclidean action
\begin{equation}
S_{E}(\varphi , x)
=-\frac{1}{2}\int_{\eta_{i}}^{\eta_{f}}d\eta\varphi^{2}((\frac{dx}{d\eta})^{2}+x^{2}-\frac{1}{2}m^{2}\varphi^{2}x^{4}),
\end{equation}
where $m$ is the generally nonzero mass of a dimensionless constant scalar
field $\varphi$
and $\eta$ denotes a dimensionless Euclidean time $\eta=\int\frac{d\tau}{x}$.
The sign for the Euclidean action would be positive when we choose the usual
Wick rotation $t\rightarrow -i\tau$ (clockwise) or negative for a Wick rotation
$t\rightarrow +i\tau$ (anti-clockwise). If the 'energy' of the particles is
positive, then one should use clockwise rotation, but if it is negative the
rotation would be anti-clockwise [14].
It will be seen later on that (3.1) corresponds to the case of a massive field
conformally coupled to gravity, with $x$ playing the role of the
Robertson-Walker
scale factor. Thus, since the gravitational energy associated with the scale
factor is negative, one should rotate $t$ not to $-i\tau$, but to $+i\tau$,
such
as it is done in (3.1), and therefore the semiclassical path integral involving
action $S_{E}$, $e^{-S_{E}}$,
may be generally interpreted as a probability amplitude for quantum
tunnelling [14].
In the classical case, if the scalar field $\varphi$
is axionic, i.e. if $\varphi$ is assumed to be a hypothetical light
particle in the dynamical mechanism that solves the strong $CP$
problem of $QCD$,
then it becomes pure imaginary, i.e. $\varphi=i\varphi_{0}$, with
$\varphi_{0}$ real. In
this case, the potential becomes
\begin{equation}
V(x) = \varphi_{0}^{2}(\frac{1}{2}x^{2}+\frac{1}{4}m^{2}\varphi_{0}^{2}x^{4}),
\end{equation}
the theory has just one zero-energy vacuum, and the solution to
the classical equations of motion is $\bar{x}=0$, with Euclidean action
$S_{E}=0$.

Expanding about the classical solution we obtain the usual path integral at
the semiclassical limit [17]
\begin{equation}
<0\mid e^{-HT/\hbar}\mid 0>\sim
N[det(-\partial_{\tau}^{2}+\omega^{2})]^{-\frac{1}{2}}(1+O(\hbar)),
\end{equation}
where $N$ is a normalization constant and $\omega^{2}=V''(0)$, with $'$
denoting
differentiation with respect to $x$. The ground-state solution to the wave
equation for the operator $-\partial_{\tau}^{2}+\omega^{2}$ which corresponds
here
to a harmonic oscillator with substracted zero-point energy can be written as
\[\psi_{0}=\omega^{-1}e^{-\omega T}\sinh[\omega(\tau+\frac{T}{2})]. \]
Then, the path integral becomes constant and given by
\begin{equation}
<0\mid e^{-HT/\hbar}\mid 0>\sim
(\frac{\omega}{\pi\hbar})^{\frac{1}{2}}(1+O(\hbar)).
\end{equation}
If the zero-point energy had not been substracted, then Eqn. (3.4) would also
contain the well-known time-dependent factor $e^{-\frac{1}{2}\omega T}$ [17].

Let us now consider $\varphi^{2}$ as being the control parameter for the
nonlinear dynamic problem posed by action (3.1). All the values of
$\varphi^{2}$
corresponding to an axionic classical field will be negative and, in the
classical
case, can be continuously varied first to zero (a zero potential critical
point) and then to positive values (the field $\varphi$ has become real, no
longer axionic). The associated variation of the dynamics will represent a
typical classical bifurcation process that can finally lead to spontaneous
breakdown of a given symmetry [18]. In the semiclassical theory this generally
is no longer
possible however. Not all values of the squared field $\varphi^{2}$ are then
equally probable. For most cases, large ranges of $\varphi^{2}$-values along
the bifurcation itinerary are strongly suppressed, and hence the bifurcation
mechanism would not take place. Nevertheless, there could still be sudden
reversible quantum jumps from the negative values to the corresponding
positive values of $\varphi^{2}$. Such jumps would be equivalently expressible
as analytic continuations in $x$ to and from its imaginary values,
lasting a very short time. The system will first go from the bottom of
potential (3.2) for $\varphi^{2}<0$ to the sphaleron point of potential (see
Fig.1)
\begin{equation}
V(x)=\varphi_{0}^{2}(-\frac{1}{2}x^{2}+\frac{1}{4}m^{2}\varphi_{0}^{2}x^{4}),
\end{equation}
without changing position or energy, and then will be perturbed about the
sphaleron saddle point to fall into the broken vacua where the broken phase
condenses
for a short while, to finally redo all the way back to end up at the bottom
of potential (3.2) for $\varphi^{2}<0$ again. The whole process may be denoted
as a quantum sphaleron
transition and it is assumed to last a very short time and to occur at a very
low frequency along the large time $T$. Therefore, one can use a dilute
spahaleron approximation which is compatible with our semiclassical approach.
Thus, for large $T$, besides individual quantum sphalerons, there would be
also approximate solutions consisting of strings of widely separated quantum
sphalerons. In analogy with the instanton case [17], we shall evaluate the
functional integral by summing over all such configurations, with $n$ quantum
sphalerons centered at Euclidean times $\tau_{1}$,$\tau_{2}$,...,$\tau_{n}$.
If it were not for the small intervals containing the quantum sphalerons,
$V''$ would equal $\omega^{2}$ over the entire time axis, and hence we would
obtain the same result as in (3.4). However, the small intervals with the
sphalerons correct this expression. In the dilute sphaleron approximation,
instead of (3.4), one has
\begin{equation}
(\frac{\omega}{\pi\hbar})^{\frac{1}{2}}(-K_{sph})^{n}(1+O(\hbar)),
\end{equation}
where $K_{sph}$ is an elementary frequency associated to each quantum
sphaleron,
and the sign minus accounts for the feature that particles acquire a negative
energy below the sphaleron barrier (Fig.1). Note that the action changes sign
as one goes from potential (3.2) to potential (3.5) below such a barrier.
After
integrating over the locations of the sphaleron centers, the sum over $n$
sphalerons produces a path integral
\begin{equation}
<0\mid e^{-HT/\hbar}\mid 0>_{sph}\sim
(\frac{\omega}{\pi\hbar})^{\frac{1}{2}}e^{-K_{sph}T}(1+O(\hbar)),
\end{equation}
which is proportional to the semiclassical probability of quantum tunnelling
from $\bar{x}=0$ first to $\bar{x}_{\pm}=\pm(m\varphi_{0})^{-1}$ and then
to $\bar{x}=0$ again.

In Eqn.(3.7) we have summed over any number of sphalerons, since all the small
time intervals start and finish on the axis $x=0$, at the bottom of potential
(3.2).
Approximation (3.7) corresponds [17]
to a ground-state energy $E_{0}=\hbar K_{sph}$. Thus, the effect of
the quantum sphalerons should be the creation of an extra nonvanishing
zero-point energy which must correspond to a further level of quantization
which is over
and above that is associated with the usual second quantization of the harmonic
oscillator.

As pointed out before, an Euclidean action with the same form as (3.1) arises
in a
theory where a scalar field $\Phi$ with mass $m$ couples conformally to
Hilbert-Einstein gravity. Restricting to a Robertson-Walker metric with scale
factor $a$ and Wick rotating anti-clockwise,
the Euclidean action for this case becomes, after integrating by parts and
adding a suitable boundary term,
\begin{equation}
I=-\frac{1}{2}\int d\eta
N(\frac{\dot{\chi}^{2}}{N^{2}}+\chi^{2}-\frac{\dot{a}^{2}}{N^{2}}-a^{2}+m^{2}a^{2}\chi^{2}),
\end{equation}
where the overhead dot means differentiation with respect to the conformal time
$\eta=\int d\tau/a$, $\chi=(2\pi^{2}\sigma^{2})^{\frac{1}{2}}a\Phi$, $N$ is the
lapse function and $\sigma^{2}=2G/3\pi$. The equations of motion derived from
(3.8)
are (in the gauge $N=1$)
$\ddot{\chi}=\chi+m^{2}a^{2}\chi$ and $\ddot{a}=a-m^{2}\chi^{2}a$. We note
that these two equations transform into each other by using the ansatz
$\chi=ia$.
Invariance under such a symmetry manifests not only in the equations of motion,
but
also in the Hamiltonian action and four-momentum constraints. The need for Wick
rotating anti-clockwise becomes now unambiguous [14,19],
since all the variable terms in the action
become associated with energy contributions which are negative if
symmetry $\chi=ia$ holds.
Without loss of
generality, the equations of motion can then be written as the two formally
independent expressions $\ddot{\chi}=\chi-m^{2}\chi^{3}$ and
$\ddot{a}=a+m^{2}a^{3}$.
If $\chi=ia$, then $\Phi$ becomes a constant axionic field
$\Phi=i(2\pi^{2}\sigma^{2})^{-\frac{1}{2}}$.
Re-expressing action $I$ in terms of the field $\chi$ alone, one
can write for the Lagrangian in the gauge $N=1$
\begin{equation}
L(\varphi,a)=-(\frac{1}{2}\varphi^{2}\dot{a}^{2}+\frac{1}{2}\varphi^{2}a^{2}-\frac{1}{4}m^{2}\varphi^{4}a^{4}+\frac{1}{2}R_{0}^{2}),
\end{equation}
where $\varphi=\frac{\Phi}{m_{p}}$, $m_{p}$ is the Planck mass, and $R_{0}^{2}$
is an integration constant which has been introduced to account for the axionic
character of the constant field implied by $\chi=ia$; such an imaginary field
would require a constant additional surface term to contribute the action
integral.
Besides this constant term $\frac{1}{2}R_{0}^{2}$,
(3.9) exactly coincides with the Lagrangian in (3.1). For the axionic field
case where the symmetry
$\chi=ia$ holds, $\varphi^{2}=-\varphi_{0}^{2}$, with $\varphi_{0}$ real. Then
the solution to the equations of motion and Hamiltonian constraint is
\begin{equation}
a(\tau)=(m\varphi_{0})^{-1}[(1+2m^{2}R_{0}^{2})^{\frac{1}{2}}\cosh(2^{\frac{1}{2}}m\varphi_{0}\tau)-1]^{\frac{1}{2}}, \chi=ia(\tau),
\end{equation}
which represents an axionic nonsingular wormhole spacetime.

If we rewrite (3.9) as a Lagrangian density
$L(\Phi,a)=m_{p}^{2}L(\varphi,a)/a^{4}$,
it turns out that, although the Lagrangian density as written in the form
$L(\Phi,a)$
looks formally similar (except the last term in the potential) to that for an
isotropic and homogeneous Higgs model in Euclidean time, it however preserves
all the symmetries of the theory intact. Indeed, the field $\Phi$
is not but a simple imaginary constant
$\Phi=i(2\pi^{2}\sigma^{2})^{-\frac{1}{2}}$.
None the less, if $\Phi$ is shifted by some variable $\rho$, such that
$\Phi\rightarrow\phi=i\xi+\rho$, where
$\xi=(2\pi^{2}\sigma^{2})^{-\frac{1}{2}}$,
while keeping the same real scale factor, then
symmetry $\chi=ia$ would become broken.
In such a case, the Lagrangian density $L\equiv L(\phi,a)$ resulting from
$L(\Phi,a)$ is seen (Note that if we let $\Phi$ to be time-dependent, then
the kinetic part of the Lagrangian $L(\Phi,a)$ becomes
$-[\frac{1}{2}(\frac{\dot{a}}{a^{2}})^{2}\Phi^{2}$
$+(\frac{\dot{a}}{a^{2}})\Phi\frac{\dot{\Phi}}{a}+\frac{1}{2}(\frac{\dot{\Phi}}{a})^{2}$])
to describe a typical Higgs model in the isotropic and
homogeneous euclidean framework for a charge-$Q$ field $\phi$, a massless gauge
field $A\equiv A(\tau)=\frac{TrK}{eQ}$ (with $e$ an arbitrary gauge coupling
and
$K$ the second fundamental form),
and variable $tachyonic$ mass $\mu=\frac{1}{a}$, as now the last term in the
potential does not depend on $\phi$ and becomes thereby harmless for the
Higgs model. In principle, the Lagrangian could have been written as well
in any of the infinite number of forms, other than (3.9), which also
have both $\Phi=Const$ as a solution and the symmetry $\chi=ia$.
However, the system will spontaneously choose breaking the symmetry
from the particular Lagrangian form given by (3.9) for two reasons:
(i) because, in so doing, it attains the most stable vacuum
configuration, and (ii) because, out of the infinite number of
possible Lagrangians, it is only $L(\Phi,a)$ which is invariant
under the isotropic and homogeneous Euclidean ($t\rightarrow +i\tau$)
Abelian group $U(1)$ of transformations
\[A\rightarrow A-i\dot{\zeta}, \Phi\rightarrow\Phi e^{i\zeta},\]
since only for $L(\Phi,a)$ integration of $A$ yields $a\rightarrow
ae^{-i\zeta}$,
so that $\chi$ is also invariant under these transformations.
Note that
the equivalent rotation in scale factor to generally complex values in the
representation obtained by re-expressing $I$ in terms of $a$ alone will lead
to a breakdown of diffeomorphism invariance [20]. Clearly, all theories
satisfying $\chi=ia$ must contain an axionic surface term
$-\int d\eta NR_{0}^{2}$ which is of course invariant under such
a symmetry. Then, it is easy to show that both, the equations of
motion and the Hamiltonian constraint derived from (3.8) satisfy
solution (3.10). It is in this sense that the scalar field theory
and the theory which shows sphaleron transitions (3.9) are
equivalent, though only the latter is invariant under the
transformations of the gauge group $U(1)$.

The integration constant $R_{0}^{2}$ in (3.9) corresponds to the inclusion of
axionic
surface terms in the action integral. For a constant real scalar field, such
surface terms should
produce a generally different integration constant $(R'_{0})^{2}$ whose sign is
the opposite to
that for $R_{0}^{2}$. Therefore, the solutions to the equations of motion and
Hamiltonian constraint corresponding to a constant real scalar field
$\varphi^{2}=\varphi_{0}^{2}$ are
\begin{equation}
a_{\pm}(\tau)=(m\varphi_{0})^{-1}[1\pm(1-2m^{2}(R'_{0})^{2})^{\frac{1}{2}}\cosh(2^{\frac{1}{2}}m\varphi_{0}\tau)]^{\frac{1}{2}}.
\end{equation}
Solutions $a_{+}$ and $a_{-}$ lie in the two disconnected, classically allowed
regions for which the potential is negative. It has been shown [21] that
(3.11) represents a wormhole with doubly connected inner region whose
quantum state should be given by a statistical density matrix.

The real classical solution corresponding to the equilibrium minimum of the
potential for axionic fields is $\bar{a}=0$, and that for real fields are
$\bar{a}_{\pm}=\pm(m\varphi_{0})^{-1}$. Therefore, relative to the simply
connected inner topology implied by (3.10), the doubly-connected topology
implied
by (3.11) represents an actual pitchfork bifurcation for $R_{0}=R_{0}'$.

We can see now why not all values of $\varphi_{0}$ are allowed semiclassically.
Although one actually could also gauge the imaginary part of $\phi$ to any
value other than $\xi$,
since the action $I$ depends on $\varphi_{0}^{-2}$ through solution (3.10),
very small values of
the constant fields will make this action very large and hence the
semiclassical
probability $e^{-I}$ becomes vanishingly small. Thus, interpreting
$\varphi^{2}$ as
a control parameter it turns out that the bifurcation process cannot be reached
in a continuous, deterministic way.

Disregarding the constant term of the potential in (3.9) to make this potential
vanish
at its minima, we obtain the same situation as in Fig.1, with $\omega\sim 1$ in
Planck units. If it were not for the short intervals where sphalerons are
quantically induced, the contribution of the path integral $<0\mid
e^{-HT/\hbar}\mid 0>$
to the quantum state of the system would simply be a constant factor. This
would correspond to a pure quantum state representable by a probability
functional
factorizable in a product of wave functions.
However,
if such sphaleron transitions (essentially consisting of rotations of the
metric to
generally complex values)
are taken into account, then the contribution of
the path integral $<0\mid e^{-HT/\hbar}\mid 0>_{sph}$ will introduce a
time-dependent
factor like (3.7), with $\omega\sim 1$, in the full quantum state. Now, since
time separation between the initial and final states cannot be known, one
should
integrate over $T$ and the full quantum state becomes (here and hereafter we
omit for the sake of simplicity the small term depending on $O(\hbar)$)
\begin{equation}
\frac{1}{\tau_{p}}\int_{0}^{\infty}dT\Psi[a]\Psi[a']e^{-K_{sph}T},
\end{equation}
where $\tau_{p}$ is the Planck time and
the $\Psi '$s are the wave functions for the initial and final states.
{}From the Euclidean action (3.8), after applying $\chi=ia$, we obtain a
Wheeler
DeWitt equation
\begin{equation}
H_{WDW}\Psi=(\frac{\partial^{2}}{\partial a^{2}}+V(a,m,R_{0}))\Psi = 0,
\end{equation}
where the potential $V(a,m,R_{0})=R_{0}^{2}-a^{2}-\frac{1}{2}m^{2}a^{4}$.
In harmonic approximation ($m\rightarrow 0$), we obtain for the quantum state
\begin{equation}
\Psi_{j}(a)=H_{j}(a)e^{-\frac{1}{2}a^{2}};
\end{equation}
i.e., after symmetry $\chi=ia$, the wormhole wave function is given by either
$\Psi(a)$, or by the anologous function $\Psi(\chi)$, but $not$ by their
product.

The contribution from the strings of single quantum sphalerons bouncing at the
vacua of potential (3.5) are not the only objects that should contribute the
path integral. Once the system has returned to the top of this potential after
one transition is done, it may yet undergo further sphaleron-asisted
bifurcations,
before completing transition to the bottom of potential (3.2) within the given
small time interval.
This would lead to a differentiation of the transition strings:
there will be strings formed by $n_{1}$ single transitions consisting of
just one quantum sphaleron, strings formed by $n_{2}$ single transitions
consisting of two quantum sphalerons and, in general, strings formed by $n_{j}$
single transitions consisting of $j$ quantum sphalerons. So, considering
the contributions of all such transition strings leads to the maximal analytic
extension of the wormhole manifold. In what follows we shall show how such
contributions may enter the density matrix for wormholes. We shall use
first-order
time-dependent perturbation theory in our Euclidean framework in order to
independently derive an expression for the density matrix in terms of states
(3.14). Let us first
define the factorizable pure-state density matrix elements as
$\rho_{j}^{(0)}=\Psi_{j}(a)\Psi_{j}(a')$,
so that $H_{WDW}\rho_{j}^{(0)}=0$. Considering the wormhole to be off shell
one can introduce an energy perturbation $H(\tau)$. We have then
\begin{equation}
(H_{WDW}+H(\tau))\rho(a,\tau)=\hbar\frac{\partial\rho(\tau)}{\partial\tau}.
\end{equation}
In the limit of very small $m$, using the current solution of the perturbed
equation $\rho(\tau)=\sum_{j}a_{j}(\tau)\rho_{j}^{(0)}$, we obtain after
multiplying for $\rho_{k}^{(0)}$ and integrating over the scale factor
\begin{equation}
\hbar
2^{2k}(k!)^{2}\pi^{2}\frac{da_{k}(\tau)}{d\tau}=\sum_{j}H_{jk}(\tau)a_{j}(\tau),
\end{equation}
where
\begin{equation}
H_{jk}(\tau)=\int\rho_{j}^{(0)}H(\tau)\rho_{k}^{(0)}dada'=
H_{jk}(0)e^{-\frac{1}{\hbar}(E_{j}+E_{k})\tau},
\end{equation}
in which $E_{j}$ and $E_{k}$ are the energies of, respectively, the $j$th and
$k$th harmonic-oscillator levels.
At least for the nonsupersymmetric case under study, the wormhole oscillators
have no zero-point energy. Thus,
choosing $E_{k}=0$ and $H(\tau)=\epsilon_{p}f(\tau)$, so that
$H_{jk}(0)=2^{2j}(j!)^{2}\pi^{2}\epsilon_{p}$,
where $\epsilon_{p}$ is the Planck-scale energy, we obtain in first order
\begin{equation}
\hbar\frac{da_{j}^{(1)}}{d\tau}=\epsilon_{p}e^{-\frac{1}{\hbar}E_{j}\tau},
\end{equation}
For the most probable wormholes
with the Planck size, $E_{j}=j\epsilon_{p}$ in the harmonic approximation.
Hence,
\begin{equation}
\rho(\tau)=\frac{1}{\tau_{p}}\sum_{j=1}^{\infty}\int
d\tau\Psi_{j}[a]\Psi_{j}[a']e^{-\frac{j\epsilon_{p}\tau}{\hbar}}.
\end{equation}
The integrand in (3.19) will correspond, for large $\tau$, to semiclassical
approximation of the
path integral for our Euclidean framework. The lower limit of index $j$ is
taken
to be 1 because index $k$ stands for the zero-energy ground state and $j\neq
k$.
Thus, the argument [17] that the path integral in semiclassical approximation
must equal the first term of its expansion in energy eigenvalues should
reflect here in such a way that (3.12) will be the same as (3.19) for $j=1$,
and hence $\hbar K_{sph}=\epsilon_{p}$.
It follows then ($\tau=T$)
\begin{equation}
\rho(T)=\frac{1}{\tau_{p}}\sum_{j=1}^{\infty}\int_{0}^{\infty}dT\Psi_{j}[a]\Psi_{j}[a']e^{-jK_{sph}T},
\end{equation}
which is the expression that one should obtain by summing up all contributions
from the different strings formed by $n_{j}$ single elementary quantum
sphalerons
with $K_{sph}=\epsilon_{p}$ in the harmonic approximation.

As discussed in Section 2, density matrices defined as the propagator
(2.4) [22] are divergent [23] and generally not positive definite [13].
This is due to the fact that integration over $\tau$ leaves a factor
$\epsilon_{mn}^{-1}$ in the summation over $m$ and $n$, producing a
divergent term when $\epsilon_{mn}=0$. Now, since $\Psi[a]$ does not
depend on $T$, $K_{sph}>0$, and the lower bound for index $j$ is unity,
to the extend that $\Psi$ be convergent, the state (3.20) will give an
also convergent, positive definite density matrix for nonsimply
connected wormholes. On the other hand, in the limit where either
$\hbar\rightarrow 0$ or $G\rightarrow 0$, or both, taken after
integrating over $T$, the path integral (3.20) vanishes, so
leaving no wormhole state. This implies (3.20) to be a quantum
state with no classical counterpart, i.e. the density matrix
of wormholes is a {\it nonclassical} state similar to those occurring
in quantum optics. Moreover, $K_{sph}$ corresponds to a kind of
zero-point energy which does not take place in usual quantum
field theories and, therefore, even though the sphaleron
contribution has been worked out here in a semiclassical approximation,
it must still represent an extra level of quantization above that
is contained in the quantum theories. The vanishing of the path
integral in the nongravitational limit
suggests that the extra level of quantization
introduced by quantum sphalerons can only appear when quantum-gravity effects
are considered, or in other words, quantum gravity involves an extra level of
quantization which is over and above that is contained in nongravitational
quantum theory. We see that the introduction of this extra quantization
produces
three main effects: (i) it converts pure states in mixed states describable by
a density matrix, (ii) it breaks energy degeneracy (all those
quantum numbers, $n$ and $m$ in (2.5), leading to the same eigenenergy
$\epsilon_{mn}$)
by reducing the set of eigenenergies $\epsilon_{mn}$ to the subset
$\epsilon_{j}=jK_{sph}$,
and (iii) it makes the density matrix finite since $K_{sph}>0$.

The extra degree of quantization involved in (3.20) can be thought of as being
originated from a discretization of time separation [24] so that
$\mid\tau\mid\geq\tau_{p}$, where $\tau_{p}$ is the Planck time, at the times
a quantum sphaleron occurs. In fact, work done in quantum gravity [25-27]
predicts that it is altogether
impossible to get a spacetime resolution better than the Planck scale. If we
take Planck-sized
wormholes, then it follows [24] that,
as one approaches the time resolution
$\bigtriangleup\tau=\tau_{p}$ at any $\tau$-time point along an initially
single
wormhole tube, a topologically nontrivial bifurcation will develop at the given
point. Thus, expliciting the discrete character of $\tau$ at (to keep things
in semiclassical and dilute bifurcation approximations) widely separated points
converts any simply
connected wormhole in a multiply connected wormhole with any number of (in the
harmonic approximation) identical bifurcations.
The quantum state of the resulting system is given by (3.20) and would
be equivalent [24] to that for a single wormhole-tube with $continuous$ time
$\tau$
having contributions from all possible discrete energies $j\varepsilon_{p}$,
with $j>0$ for $\tau\equiv T>0$, $j<0$ for $\tau\equiv T<0$, and $1\leq\mid
j\mid\leq\infty$,
where $\varepsilon_{p}=K_{sph}$ denotes Planck energy for a Planck-sized
wormhole.
Discreteness in time $\tau$ can thus be transformed into discreteness in
wormhole energy.

A caveat is worth mentioning here. It has been pointed out [19] that a
rotation $t\rightarrow +i\tau$ would imply a repulsive gravitational regime.
Nevertheless, a probability functional which is factorizable as a product of
equal wave functions [10] can no longer represent the ground state if
diffeomorphism invariance is preserved, for all eigenenergies are strictly
zero [14]. However, if diffeomorphism invariance is broken, so that a ground
state as (3.20) becomes well defined, then such a ground state would be below
the barrier for $t\rightarrow +i\tau$, in a situation where the Euclidean
action is negative and corresponds therefore to a positive gravitational
constant and hence to attractive gravity.

A key feature generally distinguishing
density matrices from wave functions in the context of quantum cosmology
is that, whereas the Wheeler DeWitt operator,
$H_{WDW}$, annihilates the state $\Psi$, that is no longer the case when it
acts on $\rho$, i.e. $H_{WDW}\rho$ $= \epsilon_{mn}$ $\neq 0$, meaning that the
total energy of the baby universe sector
is no longer zero (i.e. a virtual baby universe is no longer a closed universe
[28]). In
that case, gravitational flow lines from the baby universes would in fact
connect
them to the asymptotic region [29].
This effective topological connection between
the baby universe spacetime manifold and the main large manifold should affect
the whole wormhole four-manifold $M$, so that,
even if this was originally conventionally divided into two parts, $M_{+}$ and
$M_{-}$, by the three-surfaces, the resulting $effective$ wormhole
four-manifold
would become connected through the large regions, making thereby the resulting
density matrix non-factorizable as a product of wave functions $\Psi_{+}$ and
$\Psi_{-}$ [13,29]. Or in other words, an observer who wants to measure the
metric
and matter fields on any set of connected three-surfaces in the inner boundary
must approach the Planck-scale resolution and hence produce, by act of
measurement, a corresponding set of bifurcations that will ultimately make it
impossible to divide the four-manifold in two separated parts, and so forth.

\section{WORMHOLES AND TIME SYMMETRY}
\setcounter{equation}{0}

The trace of the second fundamental form, $K$,
or some monotonic function of it, may be taken to make an
extrinsic time notion in cosmology [3,30]. In order to investigate the
time-symmetry
properties of the quantum state of wormholes as given by a density matrix,
one can replace the dependence
of the functional (2.2) on $h^{\frac{1}{2}}$, the square root of the
determinant
of the three-metric, on $S$ and $S'$, by their respective conjugate momenta,
$K$ and $K'$, using the Laplace transform
\[\rho[\tilde{h}_{ij},\phi_{0},K;\tilde{h'}_{ij},\phi '_{0},K']\]
\begin{equation}=
N\int_{0}^{\infty}d[h^{\frac{1}{2}}]d[h'^{\frac{1}{2}}]e^{-\frac{1}{12\pi
G}(\int Kh^{\frac{1}{2}}d^{3}x + \int
K'h'^{\frac{1}{2}}d^{3}x')}\rho[h_{ij},\phi_{0};h'_{ij},\phi
'_{0}],\end{equation}
where $N$ is some normalization constant, $\tilde{h}_{ij}$ and
$\tilde{h'}_{ij}$ are
the three-metrics on $S$ and $S'$ up to the conformal factors
$h^{-\frac{1}{3}}$
and $h'^{-\frac{1}{3}}$, respectively, and we have restricted to the simplest
case where $S$ and its copy $S'$ are both connected three-surfaces. The density
matrix $\rho[h_{ij},\phi_{0};h'_{ij},\phi '_{0}]$ is given by (2.2) and can, in
turn, be obtained from (4.1) by the inverse Laplace transform
\[\rho[h_{ij},\phi_{0};h'_{ij},\phi '_{0}]\]
\begin{equation}=-\frac{i}{24\pi}\int_{\Lambda}d[K]\int_{\Lambda
'}d[K']e^{\frac{1}{12\pi G}(\int Kh^{\frac{1}{2}}d^{3}x + \int
K'h'^{\frac{1}{2}}d^{3}x')}\rho[\tilde{h}_{ij},\phi_{0},K;\tilde{h'}_{ij},\phi
'_{0},K'],\end{equation}
where, to ensure positivity of $h^{\frac{1}{2}}$ and $h'^{\frac{1}{2}}$, the
contours of integrations $\Lambda$ and $\Lambda '$ must run from $-i\infty$ to
$+i\infty$ to the right of any singularity of (4.1) in the complex planes $K$
and $K'$.

Yet, the Euclidean action $I[g_{\mu\nu},\phi]$ is not bounded from below and
above. Therefore,
in order to make the path integral convergent, one should deform the contour
of integration in (2.2) from Euclidean to complex metrics [31], with the
arguments of the functional defined on the inner boundary $\partial_{2}M$
and the asymptotically flat boundary $\partial_{1}M$.
The need for an inner
boundary for wormhole manifolds would incorporate the quantum-gravity idea [26]
that there exists a finite maximum spacetime resolution limit which restricted
the density of spacetime foliation to be always finite for finite
three-geometries.
Then for a conformal transformation $\bar{g}_{\mu\nu} = \Omega^{2}g_{\mu\nu}$,
$\bar{\phi} = \Omega^{-1}\phi$, where $\Omega = 1 + iy$, $y$ should be
subjected to the
boundary conditions, $y \rightarrow 0$, as $M \rightarrow \partial_{1}M$, and
$y \rightarrow y_{0}\neq 0$, as $M \rightarrow \partial_{2}M$. This choice,
which leads to a complexified fixed $h_{ij}$ and hence to a complexified second
fundamental form $K_{ij}$, so as a complexified matter field, on the inner
boundary, reflects the feature that, for a wormhole
manifold, one should replace a point at which the three-geometry degenerates
by a minimal nonzero three-surface $\partial_{2}M$ at the Planck scale [25].
One
would allow the value of $y$ to be strictly zero only at the classical
boundary $\partial_{1}M$ where the three-geometry becomes infinite, at least
in the dilute wormhole approximation [9,10]. Actually, the insertion of the
Planck three-sphere [25] is equivalent to quantizing the conformal factor so
that
conformal fluctuations would result in a "zero-point" length $l_{p}$,
according to $a^{2}=a_{class}^{2}+l_{p}^{2}$ and $\bigtriangleup a\geq l_{p}$,
where $\bigtriangleup a$ denotes quantum uncertainty in the value of $a$ [27].
But if this uncertainty holds then the baby universes can no longer be
topologically closed [28], and this will amount to a nonzero imaginary part of
the conformal factor
also on the inner boundary. A more technical reason for the need of complex
metrics on the inner boundary of multiply connected wormholes stems from
the discussion of section 3 where it was seen that a nonfactorizable
probability
functional can only be obtained if we analytically continue the dynamical
equations in the metric to generally complex values. No multiple
connectedness of the inner manifold can be preserved otherwise. The shape
of the potentials for the pure and mixed states tend to coincide as
$a\rightarrow \infty$ and can therefore
be both defined in terms of a real scale factor only asymptotically. It follows
then that the metric for multiply connected wormholes should be complexified
everywhere in the manifold, except asymptotically.
Making $y$ different
from zero also on the inner boundary is associated with the spontaneous
symmetry breaking process whose broken vacuum phase would fix the
value of $y_{0}$; the symmetry is restored asymptotically while the Higgs
mass is stored in the baby universe sector.

Because the Laplace transform (4.1) is holomorphic for
Re($K$) $> 0$ [3], one can analytically continue
$\rho[\tilde{h}_{ij},\phi_{0},K]$
in $K$ and $K'$ to the Lorentzian values
$K_{L} = iK$ and $K'_{L} = iK'$. Now, since under conformal transformation of
the
metric $\bar{g}_{\mu\nu} = \Omega^{2}g_{\mu\nu}$, we have $\bar{K} =
\Omega^{-1}K$
and $\bar{K'} = \Omega^{-1}K'$, it follows
\begin{equation}\bar{K}_{L} = \frac{(1\mp iy_{0})K_{L}}{1+y_{0}^{2}},
\end{equation}
(and likewise for $\bar{K'}_{L}$) where the upper sign corresponds to complex
metrics, and the lower sign, to
complex conjugate metrics.
The choice of time orientation on the inner
three-manifold must be based on a causal analysis requiring the definition of
suitable covering manifold [32], i.e. it is required that not more than one of
the possible several disjoint components of the three-boundary is joined at
$S$ to its respective copy $S'$.

Using a conformal metric, after analytically continuing (4.1) in $K$ and $K'$
to their Lorentzian values, we obtain from (4.3)
\[\rho^{(c)}[\tilde{h}_{ij},K_{L},\phi_{0};\tilde{h'}_{ij},K'_{L},\phi
'_{0}]=N\int_{C^{(c)}}d[(\bar{h}^{(c)})^{\frac{1}{2}}]d[(\bar{h'}^{(c)})^{\frac{1}{2}}]\]
\begin{equation}\times e^{\frac{1}{12\pi G}[\frac{1}{1+y_{0}^{2}}\int
d^{3}x(i+y_{0})K_{L}(\bar{h}^{(c)})^{\frac{1}{2}}+\frac{1}{1+(y'_{0})^{2}}\int
d^{3}x'(i+y'_{0})K'_{L}((\bar{h'})^{(c)})^{\frac{1}{2}}]}\rho^{(c)}[\bar{h}_{ij},\bar{\phi}_{0};
\bar{h'}_{ij},\bar{\phi '}_{0}]\end{equation}

\[\rho^{(cc)}[\tilde{h}_{ij},K_{L},\phi_{0};\tilde{h'}_{ij},K'_{L},\phi
'_{0}]=N\int_{C^{(cc)}}d[(\bar{h}^{(cc)})^{\frac{1}{2}}]d[(\bar{h'}^{(cc)})^{\frac{1}{2}}]\]
\begin{equation}
\times e^{\frac{1}{12\pi G}[\frac{1}{1+y_{0}^{2}}\int
d^{3}x(i-y_{0})K_{L}(\bar{h}^{(cc)})^{\frac{1}{2}}+\frac{1}{1+(y'_{0})^{2}}\int
d^{3}x'(i-y'_{0})K'_{L}(\bar{h'}^{(cc)})^{\frac{1}{2}}]}\rho^{(cc)}[\bar{h}_{ij},\bar{\phi}_{0};
\bar{h'}_{ij},\bar{\phi '}_{0}],
\end{equation}
where the superscripts $(c)$ and $(cc)$ mean quantities obtained when,
respectively,
only complex metrics and complex conjugate metrics are used. Integrations over
contours $C^{(c)}$ and $C^{(cc)}$ respectively extend
$(\bar{h}^{(c)})^{\frac{1}{2}}$
and the orientation reverse of $(\bar{h'}^{(c)})^{\frac{1}{2}}$ from 0 to
$\infty$ in
the complex plane for contour $C^{(c)}$, and $(\bar{h}^{(cc)})^{\frac{1}{2}}$
and the orientation reverse of $(\bar{h'}^{(cc)})^{\frac{1}{2}}$ also from 0 to
$\infty$,
but in the complex conjugate plane for contour $C^{(cc)}$. The lower
integration
limits have been chosen to be zero to allow for a total quantum freedom both
for the three-geometry
and slicing. Note that the values of the three-metrics, scalar fields and trace
of the second fundamental form induced on $\partial_{2}M$ are all still
complex.

Applying complex conjugation ($\ast$) and the operation of $K_{L}$-reversal to
states (4.4) and (4.5), we obtain
\begin{equation}\rho^{(c)}[\tilde{h}_{ij},-K_{L},\phi_{0};\tilde{h'}_{ij},-K'_{L},\phi '_{0}]^{\ast}=\rho^{(cc)}[\tilde{h}_{ij},K_{L},\phi_{0};\tilde{h'}_{ij},K'_{L},\phi '_{0}] \end{equation}

\begin{equation}\rho^{(cc)}[\tilde{h}_{ij},-K_{L},\phi_{0};\tilde{h'}_{ij},-K'_{L},\phi '_{0}]^{\ast}=\rho^{(c)}[\tilde{h}_{ij},K_{L},\phi_{0};\tilde{h'}_{ij},K'_{L},\phi '_{0}]. \end{equation}

It follows that if the full density matrix contained equal contributions from
metrics with a complex action and from metrics with a complex conjugate action,
then that density matrix would be $T$ symmetric. However, fixing prescribed
complex values for the boundary metric and scalar field makes the density
matrix and its Laplace transform holomorphic functions, i.e. any contribution
from metrics with complex conjugate action to the path integral in (4.4), or
from metrics with a complex action to the path integral in (4.5) must be ruled
out. It follows then that, fixing a complex value for the boundary metric and
the boundary scalar field, contributions from complex conjugate four-metrics
can never enter the path integral or its Laplace transform, and likewise,
fixing
the corresponding complex conjugate values for the boundary arguments will
prevent any four complex-metric to contribute the path integral or its Laplace
transform.

Since the Wheeler-DeWitt operator annihilates any pure-state wave function of
wormholes, for such states we must sum over both complex and complex conjugate
four-metrics which induce the prescribed boundary real metric. Thus, by
analytically continuing in the metric
the single
wave functions for wormholes, $\Psi_{+}$ and $\Psi_{-}$, should respectively
map into the two distinct
density matrices $\rho^{(c)}$ and $\rho^{(cc)}$, and their Laplace transform
respectively into
(4.4) and (4.5). We have,
\begin{equation}\rho^{(c)}[\tilde{h}_{ij},K_{L},\phi_{0};\tilde{h'}_{ij},K'_{L},\phi '_{0}] \neq \rho^{(c)\ast}[\tilde{h}_{ij},-K_{L},\phi_{0};\tilde{h'}_{ij},-K'_{L},\phi '_{0}],\end{equation}

\begin{equation}\rho^{(cc)}[\tilde{h}_{ij},K_{L},\phi_{0};\tilde{h'}_{ij},K'_{L},\phi '_{0}] \neq \rho^{(cc)\ast}[\tilde{h}_{ij},-K_{L},\phi_{0};\tilde{h'}_{ij},-K'_{L},\phi '_{0}],\end{equation}
and
\begin{equation}\Psi[\tilde{h}_{ij},K_{L},\phi_{0}]=\Psi^{\ast}[\tilde{h}_{ij},-K_{L},\phi_{0}].\end{equation}

If we interpret $\rho^{(c)}$ as a functional relative to observers in one
of the two asymptotic regions, and $\rho^{(cc)}$ as a functional relative
to observers in the other asymptotic region, then
Eqns. (4.8) and (4.9) are a statement of $T$ noninvariance for the quantum
state
of wormholes for independent observers in the regions which are connected
by the wormhole. This would imply that, for an observer in one large region,
the probability for creating baby universes is not the same as that for
destroying
them. The wormhole state representation $\Psi$ should be time-symmetric
however; for (4.10)
contains equal contributions from metrics with a complex action and from
metrics
with a complex conjugate action so that the wave function is real.

Moreover, if following Hawking [3], we replace the real scalar field by a
complex
one, and introduce a triad of covectors on the three-surface, so as a fermion
field,
one can show that, whereas a pure quantum state of wormholes also is $CT$ and
$CTP$
invariant, wormholes describable by a density matrix must be noninvariant under
these operations.

There is yet an important property [33] which could modify the extent at
which the rhs and lhs of (4.8) and (4.9) are different. It is that the complex
conjugate, $\bar{\phi}$, of a scalar field $\phi$ may be replaced by
an independent scalar field $\tilde{\phi}$ which is not related to
$\phi$ by any conjugation. The same applies to spinor fields. However,
since such a property would not obviously affect the complex values
of the three-metrics and extrinsic curvature tensors, even if it applied
to the time-symmetry treatment [3], the above analysis would only be
modified in such a way that inequalities (4.8) and (4.9) will still
hold.

\section{THE TOPOLOGICAL ARROW OF TIME}
\setcounter{equation}{0}

The demand of locality on quantum fields in the asymptotic regions is made
possible
by the $T$ invariance of state $\Psi$, and implies that the effective
interaction
Hamiltonian density in Minkowski space $H_{i}A_{i}$ must commute for
nonzero spacelike separations. Here, the $H_{i}'$s are scalar-field interaction
operators terms, the $A_{i}'$s are baby universe operators which are given in
terms
of Fock operators [21,34], and $i$ is a discrete index characterizing baby
universes.
We note now that the quantum states $\rho^{(c)}$
($\rho^{(cc)}$): (i) are Lorentz invariant because wormhole spacetime is
asymptotically flat (Notice that baby universes being branched off and in
through
wormholes whose quantum state is not factorizable as a product of wave
functions
are always gravitationally connected to the asymptotic regions),
and (ii) have positive definite energy which is given by $\hbar K_{sph}$.
Hence, for states $\rho^{(c)}$ and $\rho^{(cc)}$, the lack of $T$, $CT$ and
$CTP$ invariance
must be addressed [35] to a breakdown of local causality in the baby universe
sector,
\begin{equation}[A_{i}(x),A_{j}(y)] \neq 0 , (x-y)^{2} < 0.\end{equation}
{}From the overall demand of locality, it follows then,
\begin{equation}[H_{i}(x),H_{j}(y)] \neq 0 , (x-y)^{2} < 0, \end{equation}
which, in turn, implies that the set of all possible quantum fields in the
large region able to
interact with wormholes is not $CTP$ invariant, as far as these fields are
Lorentz
invariant and have positive energy. Clearly, although $CTP$ invariance appears
to be violated in both,
the baby universe and low-energy matter subsystems separately, overall demand
of locality will ensure $CTP$ invariance for the total, joint system because
of its Lorentz invariance and energy positiveness.

Now, as in most situations the effects of any
$C$ or $P$ noninvariance can be neglected, (5.2) actually implies noninvariance
of the quantum fields in the large regions under $T$. Since it appears that all
quantum fields in the large regions should interact with wormholes, associated
with the time-asymmetry of $\rho^{(c)}$ (or $\rho^{(cc)}$) there will be a
universal
arrow of time which could be named $topological$ because of its origin. That
topological arrow, which would indicate the direction in which the two wormhole
large
regions are created and evolve relative to each other, becomes the basic
relevant
information contained in the density matrix of wormholes which is accessible
to observers in the large regions, and makes $\rho^{(c)}$ and $\rho^{(cc)}$
and their Laplace transforms actual quantum states for wormholes by themselves.

If (5.1) and (5.2) hold, then there will be a topological arrow and also some
direction in which entropy increases in the large regions, for these equations
imply as well a loss of quantum coherence of the quantum fields at low energy
[21,34], and this means a thermodynamical arrow. This universal decoherence
process will furthermore induce the creation of bunched or antibunched particle
states [36]. Bunched particle states would be expected as a property of
incoherent matter whose particles are detected not at randomly
distributed times but in the form of clusters or bunches, with a
second-order coherence function larger than unity. On the other
hand, there could also exist quantum particle states which do not
admit any classical description, showing an opposite behaviour [37].
For such antibunched states the probability of detecting a
coincident pair of particles is less than that from a fully
coherent field with a random Poisson particle distribution: it
would be as if the particle detection events had some sort
of mutual repulsion.

For zero or very small particle number densities, the second-order
coherence function for matter fields can be approximated as [30]
\begin{equation}g^{(2)}_{0} \simeq
\exp(-7\sinh[2(\frac{-(x-y)^{2}}{l_{p}^{2}})^{\frac{1}{2}}]t).\end{equation}
At any time $t>0$, there will appear antibunched states characterized by {\it
nonclassical}
$g^{(2)}_{0}<1$ for given spacelike separations $(x-y)^{2}<0$, and this means
increasing time separations between detection events, just as if the sources
were radially receding isotropically from any observer.

For large particle densities, the second-order coherence function becomes [30]
\begin{equation}g^{(2)}_{n} \simeq
\frac{1}{2}(1+\exp(2\sinh[2(\frac{-(x-y)^{2}}{l_{p}^{2}})^{\frac{1}{2}}]t)).
\end{equation}
It appears that in this case one obtains bunched particles states. Such a
behaviour is expected for classical states which evolve from being pure
initially
into thermal distribution. Here, time separations between detection events tend
to decrease because of particle bunching. Therefore, the topological arrow
could
also imply directions in which the whole large regions expanded or contracted,
that is a cosmological arrow, and direction in which some restricted
fluctuating
regions shrank or expanded, that is directions in which galaxies, galaxy
clusters,
etc, and voids could develop.
Thus, it may be thought that
the strong connection [3,38] between the cosmological and thermodynamical
arrows and
the electromagnetic and psychological arrows could make the
suggested topological arrow the actual cause of all known time directions.

Work on wormholes in the realm of Euclidean quantum gravity is certainly not
free from serious problems. Therefore, although we have alleviated our
framework of some of these problems, the part of this paper that refers
to this subject still has a rather speculative nature.
\vspace{0.5cm}

$Acknowledgements$: The author wants to thank Professor Robert Geroch for
supplying valuable information. This work was supported by an Accion Especial
of C.S.I.C., and a CAICYT Project N§ 91-0052.
\pagebreak

REFERENCES

[1] S.W. Hawking, Phys. Rev. D13, 191 (1976); D14, 2460 (1976).

[2] R. Penrose, in {\it General Relativity: An Einstein Centenary Survey}, ed.
S.W. Hawking
and W. Israel (Cambridge Univ. Press, Cambridge, UK, 1979); in {\it Quantum
Gravity 2:
A Second Oxford Symposium}, ed. C.J. Isham, R. Penrose and D.W. Sciama
(Clarendon
Press, Oxford, 1981).

[3] S.W. Hawking, Phys. Rev. D32, 2489 (1985); in {\it 300 Years of
Gravitation}, ed.
S.W. Hawking and W. Israel (Cambridge Univ. Press, Cambridge, UK, 1987).

[4] R. Penrose, in {\it Proceedings of the First Marcel Grassmann Meeting on
General Relativity}
(ICTP Trieste), ed. R. Ruffini (North-Holland, Amsterdam, 1977); {\it The
Emperor's
New Mind} (Oxford Univ. Press, Oxford, 1989).

[5] See, for example, K.V. Kuchar, Phys. Rev. D43, 3332 (1991);
D45, 4443 (1992); W.G. Unruh, in {\it Gravitation}, eds. R. Mann and
P. Wesson (World Scientific, Singapore, 1991); C. Rovelli, Class.
Quant. Grav. 10, 1549; 1567 (1993).

[6] J.B. Hartle and S.W. Hawking, Phys. Rev. D28, 2960 (1983).

[7] S.W. Hawking, in {\it Astrophysical Cosmology. Proceedings of the Study
Week on
Cosmology and Fundamental Physics}, eds. H.A. Bruck, G.V. Coyne and M.S.
Longair,
Pontificia Academiae Scientiarum: Vatican City, 1982.

[8] A. Vilenkin, Phys. Rev. D33, 3560 (1986).

[9] S.W. Hawking, Mod. Phys. Lett. A5, 145; 453 (1990); G.W. Gibbons and
S.W. Hawking, {\it Euclidean Quantum Gravity} (World Scientific, Singapore,
1993); J. Twamley and D.N. Page, Nucl. Phys. B378, 247 (1992).

[10] S.W. Hawking, Phys. Lett. B195, 337 (1987)  ; Phys. Rev. D37, 904 (1988).

[11] S.W. Hawking and D.N. Page, Phys. Rev. D42, 2655 (1990).

[12] C. Teitelboim, Phys. Rev. D28, 297 (1983).

[13] P.F. Gonz\'alez-D\'{\i}az,
Nucl. Phys. B351, 767 (1991).

[14] A.D. Linde, {\it Inflation and Quantum Cosmology} (Academic Press, Boston,
1990); Phys. Scripta T36, 30 (1991).

[15] A.A. Belavin, A.M. Polyakov, A.S. Schwarz and Yu.S. Tyupkin, Phys. Lett.
B59, 85 (1975); G.'t Hooft, Phys. Rev. D14, 3432 (1976).

[16] N.S. Manton, Phys. Rev. D28, 2019 (1983); F.R. Klinkhamer and N.S. Manton,
Phys. Rev. D30, 2212 (1984).

[17] S. Coleman, The Uses of Instantons, in {\it The Whys of Subnuclear
Physics},
ed. A. Zichichi (Plenum Press, New York, 1979).

[18] P.F. Gonz\'alez-D\'{\i}az, Phys. Lett. B307, 362 (1993).

[19] For a discussion on the sign of Wick rotation in quantum gravity see [14]
and also the contributions by A.D. Linde and S.W. Hawking in {\it 300 Years
of Gravitation}, eds. S.W. Hawking and W. Israel (Cambridge Univ. Press,
Cambridge, 1987).

[20] A mechanism for spontaneous breakdown of diffeomorphism invariance has
already
been considered by S.B. Giddings, Phys. Lett. B268, 17 (1991). However, the
runaway solutions arising in that mechanism are not present in our case, since
the shape of potential (3.5) corresponds to the upside down version of that
used in it.

[21] P.F. Gonz\'alez-D\'{\i}az, Phys. Rev. D45, 499 (1992).

[22] D.N. Page, in {\it Proceedings of String Theory, Quantum Cosmology
and Quantum Gravity} (World Scientific, Singapore, 1986).

[23] D.N. Page, in {\it Proceedings of the Fifth Seminar on Quantum
Gravity}, eds. M.A. Markov, V.A. Berezin and V.P. Frolov (World
Scientific, Singapore, 1991).

[24] P.F. Gonz\'alez-D\'{\i}az, Mod. Phys. Lett. A8, 1089 (1993).

[25] P.F. Gonz\'alez-D\'{\i}az, Phys. Rev. D40, 4184 (1989); D42, 3983 (1990).

[26] T. Padmanabhan, Ann. Phys. (N.Y.) 165, 38 (1985); in {\it Highlights in
Gravitation
and Cosmology}, eds B.R. Iyer et al. (Cambridge Univ. Press, Cambridge, 1988).

[27] T. Padmanabhan, Phys. Rev. D28, 745, 756 (1983).

[28] If there is a limiting length of order $l_{p}$, then the induced increase
of maximum
volume of an originally closed and isotropic universe, $\Delta
V\sim(GM)^{2}l_{p}$
(in which $M$ is the bare mass of the universe) amounts to a total energy of
the
universe which is no longer zero but of the order $\epsilon_{p}$. The meaning
of this mass is that it is a quantity of which the universe does not have
enough to
be absolutely closed (M.A. Markov, Pis'ma Z. Eksp. Teor. Fiz. 45, 61 (1987);
P.F. Gonz\'alez-D\'{\i}az,
Il Nuovo Cim. 102B,
195 (1988)).

[29] P.F. Gonz\'alez-D\'{\i}az,
in {\it Classical and Quantum Systems-Foundations
and Symmetries}, ed. H.D. Doebner (World Scientific, Singapore, 1992); in {\it
Physical
Origins of Time Asymmetry}, ed. J.J. Halliwell, J. P\'erez-Mercader, W.H. Zurek
(Cambridge Univ. Press, Cambridge, 1994).

[30] J.W. York, Phys. Rev. Lett. 28, 1082 (1972); Found. Phys. 16, 249 (1986).
The choice of an extrinsic time
may have the serious drawback that this time is generally not a spacetime
scalar
and therefore is unsuitable as a many-fingered time variable in the
corresponding
functional Schroedinger equation. A way to overcome this difficulty could be
the inclusion (K.V. Kuchar, Phys. Rev. D45, 4443 (1992)) of a scalar field $S$
such that the field equations amount to the statement that the value of $S$ be
proportional to $K$. Clearly, our Higgs model of Section 3 may implement such
a procedure as a gauge scalar field $A$ naturally arises in it which is
proportional
to $K$. Moreover, additional constraints appearing as one approaches a
stationary
point of $A$ could be ruled out if one considers the extra uncertainty in
position,
thus preventing the occurrence of further undermining of the many-fingered
time Schroedinger equation on hypersurfaces transverse to any privileged time
foliation.

[31] G.W. Gibbons, S.W. Hawking and M.J. Perry, Nucl. Phys. B138, 141 (1978).

[32] R.P. Geroch, J. Math. Phys. 8, 782 (1967).

[33] S.W. Hawking, in {\it General Relativity. An Einstein-Centenary Survey},
eds. S.W. Hawking and W. Israel (Cambridge Univ. Press, Cambridge, 1979).

[34] S. Coleman, Nucl. Phys. B307, 867 (1988).

[35] R.F. Streater and A.S. Wightman, {\it PCT, Spin and Statistics and All
That}
(Addison-Wesley Publishing Co., Redwood City, 1964).

[36] P.F. Gonz\'alez-D\'{\i}az, Phys. Lett. B293, 294 (1992).

[37] R. Loudon, in {\it Concepts of Quantum Optics}, eds. P.L. Knight
and L. Allen (Pergamon Press, Oxford, 1983).

[38] F. Hoyle and J.V. Narlikar, Proc. R. Soc. Lond. A277, 1 (1964).

\pagebreak

LEGEND FOR FIGURE

Fig. 1.- Classical bifurcation itinerary for the potential in Eqn. (3.1) for
$m=0.5$.

\end{document}